\documentclass[10pt, conference, compsocconf]{IEEEtran}
\newcommand {\apgt} {\ {\raise-.5ex\hbox{$\buildrel>\over\sim$}}\ }
\newcommand {\aplt} {\ {\raise-.5ex\hbox{$\buildrel<\over\sim$}}\ }

\ifCLASSINFOpdf
  \usepackage[pdftex]{graphicx}
\else
\fi
\begin{document}
%
\title{High performance cosmological simulations on a grid of supercomputers}


\author{\IEEEauthorblockN{Derek Groen}
\IEEEauthorblockA{Centre for Computational Science\\
University College London\\
London, United Kingdom\\
Email: d.groen@ucl.ac.uk}
\and
\IEEEauthorblockN{Steven Rieder, Simon Portegies Zwart}
\IEEEauthorblockA{Leiden Observatory\\
Leiden University\\
Leiden, the Netherlands\\
Email: rieder@strw.leidenuniv.nl,\\ 
spz@strw.leidenuniv.nl}
}



\maketitle

\begin{abstract}
We present results from our cosmological N-body simulation which consisted of
2048x2048x2048 particles and ran distributed across three supercomputers 
throughout Europe. The run, which was performed as the concluding phase of
the Gravitational Billion Body Problem DEISA project, integrated a 30 Mpc box of dark matter
using an optimized Tree/Particle Mesh N-body integrator. We ran the simulation up to the 
present day (z=0), and obtained an efficiency of about 0.93 over 2048 cores
compared to a single supercomputer run. In addition, we share our experiences 
on using multiple supercomputers for high performance computing and provide
several recommendations for future projects.
\end{abstract}


%
\IEEEpeerreviewmaketitle

\section{Introduction} 

Cosmological simulations are an efficient method to gain understanding of the
formation of large-scale structures in the Universe. Large simulations were
previously applied to model the evolution of dark matter in the Universe
\cite{2005Natur.435..629S}, and to investigate the properties of Milky-Way
sized dark matter halos \cite{Aquarius,2009ApJ...696.2115I}. However, these
simulations are computationally demanding, and are best run on large production
infrastructures. We have previously run a cosmological simulation using two
supercomputers across the globe \cite{CosmoGrid} with the GreeM integrator
\cite{2005PASJ...57..849Y,Ishiyama09}, and presented the SUSHI $N$-body
integrator \cite{sushi}, which we used to run simulations across up to four
supercomputers. The simulations we ran in the Gravitational Billion Body
Project produced over 110 TB of data, which we have used to characterize
the properties of ultra-faint dwarf galaxies~\cite{cosmogridscience}, and to
compare the halo mass function in our runs to analytical formulae for the mass
function. Among other things, we found that the halo mass function in our runs
shows good agreement with the Sheth and Tormen function
\cite{1999MNRAS.308..119S} down to $\sim 10^7$ solar mass.

Here we present the performance results of a production simulation across three
supercomputers, as well as several other runs which all use an enhanced version of
SUSHI. The production simulation ran continuously for $\sim 8$ hours, using
2048 cores in total for calculations as well as 4 additional cores for
communications. We achieved a peak performance of $3.31 \times 10^{11}$ tree
force interactions per second, a sustained performance of $2.19 \times 10^{11}$
tree force interactions per second and a wide area communication overhead of
less than 10\% overall. 

We briefly reflect on the improvements made to SUSHI for this work in Section 2, 
while we report on tests performed on a single supercomputer in Section 3. In Section
4 we describe our experiments across three supercomputers and present our performance 
results. We reflect on our experiences on using multiple supercomputers for distributed 
supercomputing simulations, and provide several recommendations for users and resource 
providers in Section 4 and present our conclusions in Section 5.


\subsection{Related work}

There are a several other projects which have run high performance computing
applications across multiple supercomputers. These include simulations of a
galaxy collision \cite{IWay}, a materials science problem \cite{paragon} as
well as an analysis application for arthropod evolution \cite{arthropod}. A
larger number of groups performed distributed computing across sites of PCs
rather than supercomputers (e.g.,
\cite{Gualandris2007,Bal20083,QCGEscience2009}). Several software tools have
been developed to facilitate high performance computing across sites of PCs
(e.g., \cite{mpich-g2,pacx-mpi,openmpi,manos,SundariHPDC}) and within volatile
computing environments \cite{RoodHPDC}. The recently launched MAPPER EU-FP7 
project \cite{mapper} seeks to run multiscale applications across a distributed 
supercomputing environment, where individual subcodes periodically exchange 
information and (in some cases) run concurrently on different supercomputing 
architectures. 

\section{Improvements to SUSHI}
Based on results of our earlier simulations and in preparation for the production 
run across three supercomputers we made several modifications to the SUSHI 
distributed $N$-body integrator. In our previous experiments a relatively large
amount of computation and communication time was spent on (non-parallelize) 
particle-mesh integration. To reduce this bottleneck we now parallelized the 
particle-mesh integration routines using the parallel FFTW2 library \cite{fftw} 
and a one dimensional slab decomposition. We also optimized the communications of 
the particle-mesh integration by introducing a scheme where sites only 
broadcast those mesh cells which have actual particle content. This optimization
reduced the size of the mesh communications by a factor roughly equal to the 
number of sites used, in the case of an equal domain distribution.

In some of the larger previous runs we also observed load imbalances if the code 
was run across two machines with different architectures, despite the presence of
a load balancing scheme. This result has led us to further optimized the load 
balancing in SUSHI, taking into account not only the force integration time, but 
also the number of particles stored on each node. In addition to these changes,
we also seized the opportunity to plug in a more recent MPWide \cite{MPWide} 
version into SUSHI. This newer version contains several optimizations to improve 
the wide area communication over networks with a high latency.

\section{Tests on a single site}

\subsection{Setup}

We performed a number of runs on the Huygens supercomputer to validate the
scalability of our new implementation, and to provide performance measurements
against which we can compare our results using multiple sites. More information
on the Huygens machine can be found in the second column of
Tab.~\ref{Tab:Superspecs}.  The initial conditions for this simulation is the
snapshot at redshift $z=0.0026$ from the CosmoGrid simulation (described in
\cite{CosmoGrid}). We also use the simulation parameters chosen for the
CosmoGrid simulation, which are summarized in Tab.~\ref{Tab:ICspecs}. Here the 
first four parameters are constants which are derived from WMAP observations 
(with a slight-roundoff) and the physical size of our simulated system is 
given by the fifth parameter (Box size). The softening in our simulation
(i.e. a length value added to reduce the intensity of close interactions) and
the sampling rate are given by the last two parameters. The sampling rate is
the ratio of particles in the simulation divided by the number of particles
sampled by the load balancing scheme.
Our simulation used a mesh size of $512^3$ cells. We ran the simulation using respectively 512 cores
and 1024 cores until $z=0.0024$, and using 2048 cores until the simulation 
completed (at $z=0$). The number of force calculations per step in the simulation
varies for different $z$ values, though these variations are neglishible for 
$z<0.01$.

\begin{table}
\centering
  \caption{Initial condition and accuracy parameters used for our simulations
with $2048^3$ particles.} 
      \begin{tabular}{ll}
      \hline
      Parameter                                       & Value \\
      \hline
      Matter density parameter ($\omega_0$)           & 0.3 \\
      Cosmological constant ($\lambda_0$)             & 0.7 \\
      Hubble constant ($H_0$)                         & 70.0 km/s/Mpc \\
      Mass fluctuation parameter ($\sigma_8$)         & 0.8 \\
      Box size                                        & $(30 {\rm Mpc})^3$ \\
      Softening for $2048^3$ particle run.            & 175 pc \\
      Sampling rate for $2048^3$ particle run.        & 20000 \\
      \hline
    \end{tabular}


  \label{Tab:ICspecs}
\end{table}

\subsection{Results}

\begin{figure}[!t]
  \includegraphics[scale=0.15]{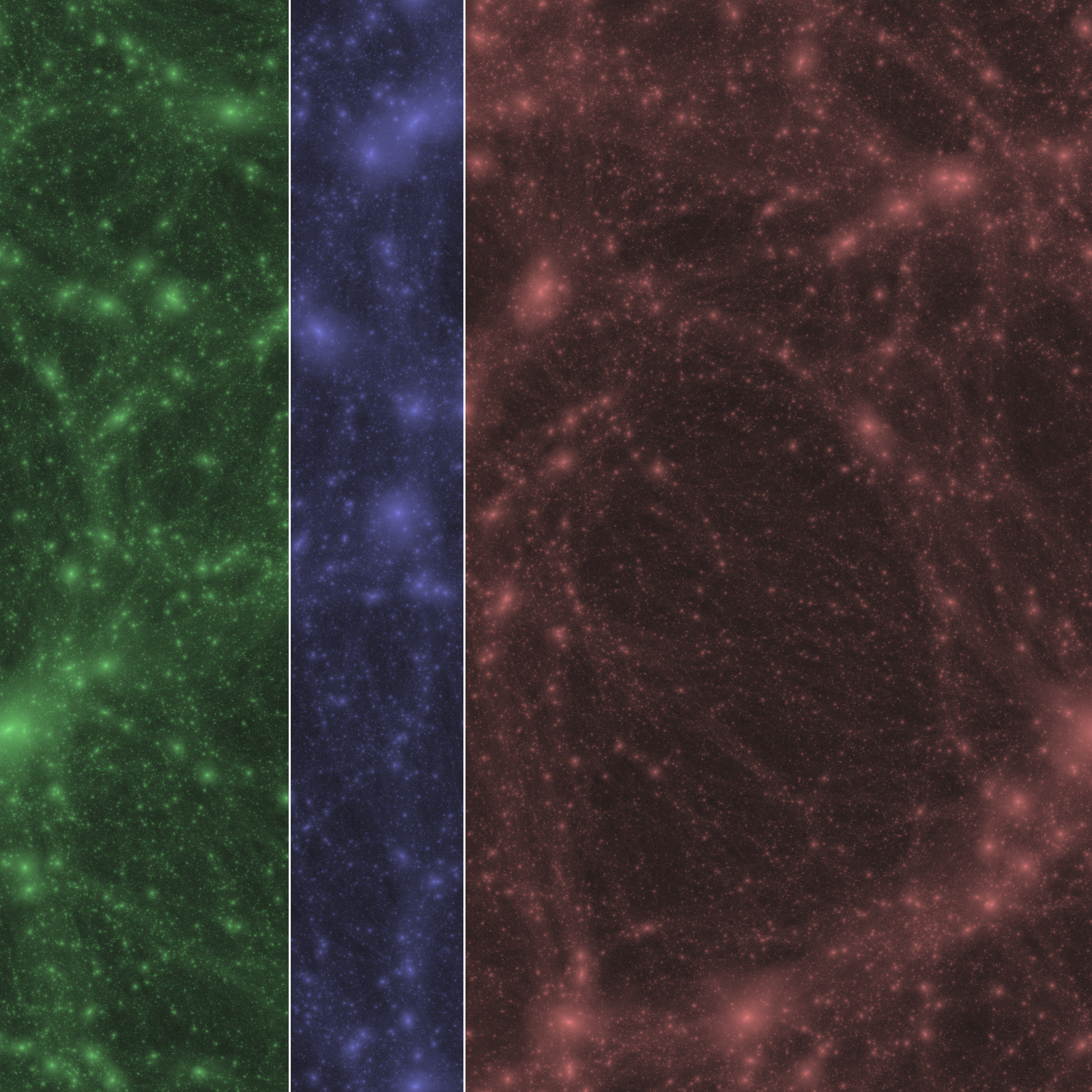}

  \caption{The final snapshot of the production simulation across three supercomputers, taken at $z=0$. The size
of the box is 30x30x30 Mpc and the contents are colored to match the particles
residing in Espoo (green, left), Edinburgh (blue, center) and Amsterdam (red, right) respectively.}

  \label{Fig:2}
\end{figure}

The performance results of our runs are shown in Tab.~\ref{Tab:1siteruns}. In
addition, the total runtime of the run using 2048 cores is given by the light
blue line in Fig.~\ref{Fig:2}.  The overall performance of the code is
dominated by calculations, with the communication overhead ranging from
$\sim$5\% for 512 cores to $\sim$10-15\% for 2048 cores. During the run using
2048 cores, several snapshots were written. This resulted in a greatly
increased execution time during two steps of the run.

\begin{table}
\centering

\caption{Overview of experiments performed with the enhanced SUSHI code on the
Huygens supercomputer. The time spent on communication is given in the fourth
column, while the total runtime is given in the fifth column. All times are
measured per step, averaged over steps 1-11. In addition we included the timing
results of the last 10 steps of the simulation running on 2048 cores (bottom
row).}

    \begin{tabular}{lllllll}

    \hline
    $N$ & $p$ & ${\theta}$ & comm. t & runtime & $z$ range & speedup\\
        &     &            & [s]     & [s]     & $\times 10^{-3}$ & \\ 
    \hline
    $2048^3$& 512 & 0.5 & 19.18 & 501.3 & 2.5-2.4 & 1\\
    $2048^3$& 1024& 0.5 & 13.96 & 258.2 & 2.5-2.4 & 1.94\\
    $2048^3$& 2048& 0.5 & 22.34 & 151.0 & 2.5-2.4 & 3.32\\ 
    $2048^3$& 2048& 0.5 & 16.22 & 143.7 & 0.1-0.0 & -\\
    \hline

    \end{tabular}


\label{Tab:1siteruns}
\end{table}

\section{Tests across three sites}

\subsection{Setup}

We performed our main run using a total of 2048 cores across three
supercomputers, which are listed in Tab.~\ref{Tab:Superspecs}. These machines
include Huygens in the Netherlands (1024 cores), Louhi in Finland (512 cores),
and HECToR in Scotland (512 cores). The sites are connected to the DEISA shared
network with either a 1Gbps interface (HECToR) or a 10Gbps interface (Huygens,
Louhi). The initial conditions and simulation parameters chosen are identical
to those of the runs using 1 supercomputer, although we use a mesh of $256^3$
cells. The use of a smaller mesh size results in a slightly higher calculation
time as tree interactions are calculated over a longer range, but a somewhat
lower time spent on intra-site communications. We configured MPWide to use 64
parallel {\tt tcp} streams per path for the wide area communication channels,
each with a {\tt tcp} buffer size set at 768 kB and packet-pacing set at 
10 MB/s maximum. We enabled some load balancing during the run, though we had to
limit the boundary moving length per step to 0.00001 of the box length due to
memory constraints on our communication nodes and the presence of dense halos
in our initial condition.

In addition to the main run, we also performed three smaller runs using the 
same code across the same three supercomputers. These include one run with 
$1024^3$ particles using 80 cores per supercomputer, and two runs with $512^3$ 
particles using 40 cores per supercomputer. These runs also used a mesh size
of $256^3$, though we did reduce the sampling rate to respectively 10000 and 
5000 for the runs with $1024^3$ and $512^3$ particles. The force softening
used for these runs were respectively 1.25kpc and 2.5kpc, and we set the 
boundary moving length limit to 0.01 of the box length. Some of the measurements 
were made using an opening angle ${\theta}$ of 0.3, rather than 0.5. Using a smaller
opening angle results a higher accuracy of the force integration on close range,
but also results in a higher force calculation and tree structure communication 
time per step.

\begin{table}
\centering
\caption{Properties of the three supercomputers used for our run. The
measured peak number of tree force interactions (in millions) per second
per core is given for each site in the bottom row.}

    \begin{tabular}{llll}

    \hline
    Name                 & Huygens    & Louhi    & HECToR    \\
    \hline
    Location             & Amsterdam  & Espoo    & Edinburgh \\
    Vendor               & IBM        & Cray     & Cray      \\
    Architecture         & Power6     & XT4      & XT4       \\
    \# of cores          & 3328       & 4048     & 12288     \\
    CPU [GHz]            & 4.7        & 2.3      & 2.3       \\
    RAM / core [GB]      & 4/8        & 1/2      & 2         \\
    force calcs. / core [Mints/s]& 185        & 256      & 250     \\
    \hline

    \end{tabular}


\label{Tab:Superspecs}
\end{table}

\subsection{Results}

The timing results of our production run are shown in Fig.~\ref{Fig:1}. Here, we also
added the wall-clock time results of the simulation run using 2048 cores on
Huygens as reference. The simulation run across three sites is only $\sim 9\%$ 
slower per step than the single-site run, despite the slightly higher force
calculation time due to the lower number of mesh cells. The peaks
in wall-clock time of the single site run are caused by the writing of
snapshots during those steps (we only wrote one snapshot at the end of the
three site run). The total wide area communication overhead of our run is $\aplt
10\%$ at about $15 s$ per step. Most of this time is required to exchange the
tree structures between sites, though the communications for the parallelized
particle-mesh require an additional $\sim 2.5 s$ per step. Despite the use of a
shared wide area network, the communication performance of our run shows very
little jitter and no large slowdowns. We provide a snapshot of the final state
of the simulation (at $z=0$), distributed across the three supercomputers, 
in Fig.~\ref{Fig:2}. 

We also provide a numerical overview of the production run performance, as well
as that of several other runs which use the new code, in
Tab.\ref{Tab:3siteruns}. The communication overhead for the runs with $512^3$
particles is less than $20\%$, while the overhead for the run with $1024^3$
particles is just $6.5\%$. The parallelization of the particle-mesh integration
and the enhanced load balancing greatly improved the performance of these runs,
especially in the case with $1024^3$ particles. Here, the communication
overhead was reduced by $\sim60\%$ and the overall runtime by more than $25\%$
compared to the previous version~\cite{sushi}.

\begin{table}
\centering
\caption{Overview of experiments performed with the enhanced SUSHI code across
all three supercomputers. All times are measured per step, averaged over 10 steps.}

    \begin{tabular}{lllllll}

    \hline
    $N$ & $p$ & ${\theta}$ & \multicolumn{2}{c}{comm. time}  & runtime & $z$ range\\
        &   &          & WAN & total &  & \\
    \hline
    $512^3$ & 120 & 0.3 & 6.925 & 7.312 & 39.70 & 11.8-10.1\\
    $512^3$ & 120 & 0.5 & 5.982 & 6.335 & 24.60 & 9.9-8.8\\
    $1024^3$& 240 & 0.3 & 12.09 & 14.04 & 214.5 & 17.0-14.9\\
    $2048^3$& 2048& 0.5 & 15.40 & 24.77 & 167.7 & 0.0026-0.0025 \\
    $2048^3$& 2048& 0.5 & 14.62 & 23.13 & 155.2 & 0.0001-0 \\
    \hline

    \end{tabular}

\label{Tab:3siteruns}
\end{table}

\section{User Experiences}

We have presented results from several cosmological simulations which run
across three supercomputers, including a production run lasting for 8 hours. In
the process of seeking a solution for wide area message passing between
supercomputers, requesting allocations, arranging network paths and preparing
for the execution of these simulations, we have learned a number of valuable
lessons.

Primarily, we found that it is structurally possible to do high performace
computing across multiple supercomputers. During the GBBP project we have run a
considerable number of large-scale simulations using two or more
supercomputers, with results improving as we were able to further enhance the
$N$-body integrator and optimize the MPWide communication library for the wide
area networks that we used. 

The cooperation of the resource providers was particularly crucial in this
project, as they enabled previously unavailable network paths and
provided us with means to initiate simulations concurrently at the different
sites. However, reserving networks and orchestrating concurrent supercomputer
runs currently does require a disproportionate amount of time and effort, which
makes performance optimization and debugging a challenging task. The effort 
required to run applications across supercomputers can be greatly reduced if 
resource providers were to adopt automated resource reservation systems for 
their supercomputers, and maintain shared high-bandwidth networking between 
sites. The persistent DEISA shared network connections helped greatly in our 
case, as we could use it at will without prior network reservations.

The software environment across different supercomputers, even within the same
distributed infrastructure, is very heterogeneous. This made it unattractive
to use existing middleware or message passing implementations to make different
sites interoperable. We chose to use a modular approach where we
connected platform-specific optimized versions of the SUSHI code with the MPWide
communication library. With MPWide being a user-space tool that requires no 
external libraries or administrative privileges, we are able to install and run 
the simulation code in the locally preferred software environments on each site 
without needing any additional (grid) middleware. We recommend adopting a similar 
modular software approach in future distributed supercomputing efforts for 
its ease of installation and optimization, at least until resource providers 
present a homogeneous and interoperable software environment for distributed 
supercomputing.

\begin{figure}[!t]
  \centering
  \includegraphics[scale=0.44]{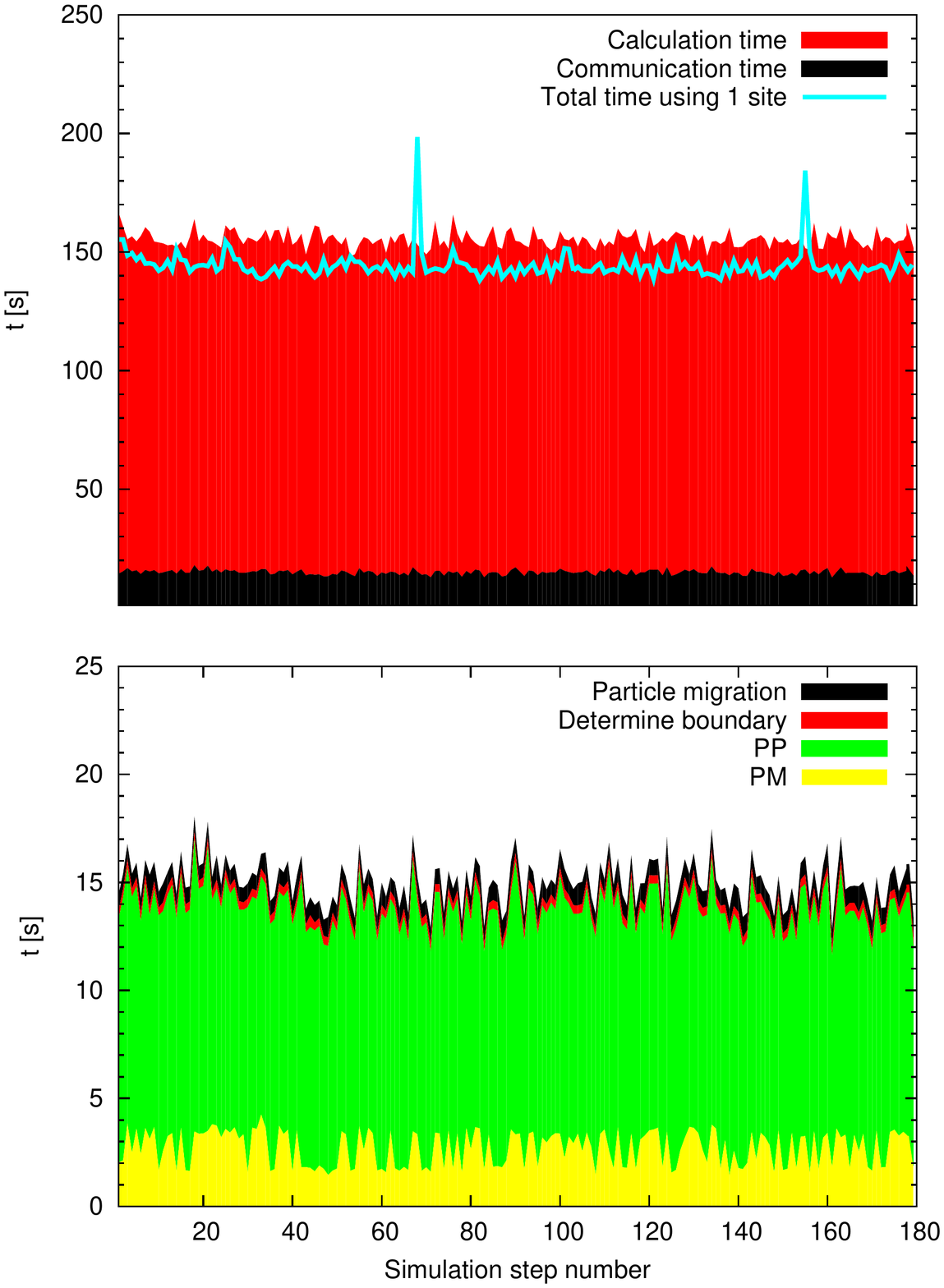}

  \caption{Performance results of the production simulation across three sites.  In the
top figure we provide the  total time spent on calculation per step in red, and
on communication per step in blue. Here, the total wall-clock time of an
identical simulation using 2048 processes only on Huygens is given by the light
blue line. Time spent on the four communication phases is given in the bottom
figure. These phases are (from top to bottom) the migration of particles
between sites, the exchanges of sample particles for determining the site
boundaries, the local essential tree exchanges (PP) and the mesh cell exchanges
(PM). See \cite{sushi} for full details on the communication routines of the
code.}

  \label{Fig:1}
\end{figure}

This paper focuses on the calculation and communication performance aspects of
a single application run across supercomputers. However, the methods presented
here can be applied for several other purposes. During this project we were
confronted with additional overhead introduced by disk I/O, as can be observed in
Figure~\ref{Fig:1}. With supercomputer disk performance and capacity improving
at a much slower rate than the compute power, the deployment of an application
across sites may help to eliminate a disk I/O performance bottleneck, though a
detailed investigation will be needed to quantify such potential benefit.
Additionally, the communication technique could be used to facilitate periodic
exchanges between different simulation codes, each of which runs on a different
site and tackles a different aspect of a complex multiscale or multiphysics 
problem.

\section{Conclusion}

Our results show that cosmological production simulations run efficiently
across supercomputers for a prolonged time. The political effort required to
arrange cross-supercomputer runs is considerable, and is an important reason
why few people have attempted to run production simulations across
supercomputers. We have shown that the added overhead of using a network of
supercomputers is rather marginal for at least one optimized production
application and that given the right (political) environment, supercomputers
can be conveniently connected to form even larger high performance computing
resources. 

\section*{Acknowledgements}
We are grateful to Jeroen B\'edorf, Juha Fagerholm, Tomoaki Ishiyama, Esko
Ker\"anen, Walter Lioen, Jun Makino, Petri Nikunen, Gavin Pringle and Joni
Virtanen for their contributions to this work.
This research is supported by the Netherlands organization for Scientific
research (NWO) grant \#639.073.803, \#643.200.503 and \#643.000.803, the
Stichting Nationale Computerfaciliteiten (project \#SH-095-08) and the MAPPER
EU-FP7 project (grant no. RI-261507). We thank the DEISA Consortium (EU FP6
project RI-031513 and FP7 project RI-222919) for support within the DEISA
Extreme Computing Initiative (GBBP project).

\bibliographystyle{IEEEtran}
\bibliography{Library}

\end{document}